\documentclass[10pt,a4paper]{article}
\usepackage{epsfig}
\usepackage{rotating}

\newcommand{\lsim}{\mathrel{\mathop{\kern 0pt \rlap
  {\raise.2ex\hbox{$<$}}}
  \lower.9ex\hbox{\kern-.190em $\sim$}}}
\newcommand{\gsim}{\mathrel{\mathop{\kern 0pt \rlap
  {\raise.2ex\hbox{$>$}}}
  \lower.9ex\hbox{\kern-.190em $\sim$}}}

\begin{document}
\textwidth=135mm
\textheight=200mm
\begin{center}

{\bfseries No role for neutrons, muons and solar neutrinos in the DAMA annual modulation results}
\vskip 5mm

R.\,Bernabei$^{a,b}$,~P.\,Belli$^{b}$,~F.\,Cappella$^{c}$,~V.\,Caracciolo$^{c}$,
\vspace{1mm}

R.\,Cerulli$^{c}$,~C.J.\,Dai$^{d}$,~A.\,d'Angelo$^{e,f}$,~S.\,d'Angelo$^{a,b}$,~A. Di Marco$^{a,b}$,
\vspace{1mm}

~H.L.\,He$^{d}$, A.\,Incicchitti$^{e,f}$,~H.H.\,Kuang$^{d}$,~X.H.\,Ma$^{d}$,~F.\,Montecchia$^{b,g}$,
\vspace{1mm}

~X.D.\,Sheng$^{d}$,~R.G.\,Wang$^{d}$ and ~Z.P.\,Ye$^{d,h}$

\vskip 5mm
{\small {\it $^{a}$ Dip. di Fisica, Universit\`a di Roma ``Tor Vergata'', I-00133  Rome, Italy}} \\
{\small {\it $^{b}$ INFN, sez. Roma ``Tor Vergata'', I-00133 Rome, Italy}} \\
{\small {\it $^{c}$ Laboratori Nazionali del Gran Sasso, I.N.F.N., Assergi, Italy}} \\
{\small {\it $^{d}$ Key Laboratory of Particle Astrophysics, Institute of High Energy Physics, Chinese Academy of Sciences, P.O. Box 918/3, 100049 Beijing, China}} \\
{\small {\it $^{e}$ Dip. di Fisica, Universit\`a di Roma ``La Sapienza'', I-00185 Rome, Italy}} \\
{\small {\it $^{f}$ INFN, sez. Roma, I-00185 Rome, Italy}} \\
{\small {\it $^{g}$ Dip. di Ingegneria Civile e Ingegneria Informatica, Universit\`a di Roma ``Tor Vergata'', I-00133  Rome, Italy}} \\
{\small {\it $^{h}$ University of Jing Gangshan, Jiangxi, China}}

\end{center}
\vskip 5mm

\centerline{\bf Abstract}

This paper summarizes in a simple and intuitive way why the neutrons, the muons and the
solar neutrinos cannot give any significant contribution to the DAMA annual modulation results. A number of these 
elements have already been presented in individual papers; they are recalled here. Afterwards,
few simple considerations are summarized 
which already demonstrate the incorrectness of the claim reported in PRL 113 (2014) 081302.

\vskip 10mm

\section{Introduction}

The DAMA/LIBRA experiment -- as the former DAMA/NaI -- at the Gran Sasso underground laboratory (LNGS) of the I.N.F.N.
is investigating the presence of the Dark Matter (DM) particles in the galactic halo by exploiting the model independent DM 
annual modulation signature. 
This DM signature is very distinctive since the effect induced by DM
particles must simultaneously satisfy all the following requirements:
i) the event rate must contain a component modulated according to a cosine function; 
ii) with period equal to one year; 
iii) with a phase roughly around June 2$^{nd}$ in case of usually adopted halo 
            models (slight variations may occur in case of presence of non thermalized 
            DM components in the halo);
iv) this modulation must be present only at low energy,
           where DM particles can induce signals; 
v) it must be present only in those events where just a single detector, 
           in a multi-detector set-up, actually ``fires'' ({\it single-hit} events), 
           since the probability that DM particles experience multiple 
           interactions is negligible; 
vi) the modulation amplitude in the region of maximal sensitivity has to be $\lsim$~7\% 
      in case of usually adopted halo distributions, but it may be significantly larger in 
      some particular scenarios.

To mimic such a signature spurious effects or side reactions should be able 
not only to account for the observed 
modulation amplitude but also to simultaneously satisfy all the requirements of the signature; 
thus, no other effect investigated so far in the field of rare processes offers
a so stringent and unambiguous signature.

In addition, let us note that neutrons, muons and solar neutrinos are not a competing background 
when the DM annual modulation signature is investigated,
since in no case they can mimic this signature (see later).
Moreover, the sensitivity of the DAMA experiments -- on the contrary of others --
is not restricted only to DM candidates giving 
rise just to nuclear recoils through elastic scatterings on target nuclei.

Let us now briefly describe the DAMA/LIBRA experiment \cite{perflibra}, recalling its model independent 
annual modulation results \cite{modlibra,modlibra2,modlibra3}. 
DAMA/LIBRA is made of 25 highly radiopure NaI(Tl) crystal scintillators, each one of 9.70 kg mass 
and size of $(10.2 \times 10.2 \times 25.4)$ cm$^3$, 
in a 5-rows 5-columns matrix. The detectors are housed in a low-radioactivity sealed 
copper box installed in the center of a passive shield
made by 10 cm of OFHC low radioactive copper, 15 cm of low radioactive lead, 1.5 mm of cadmium and about 
$10-40$ cm of polyethylene/paraffin (thickness fixed by the available space); 
moreover, about 1 m concrete (made from the Gran Sasso rock material) almost fully surrounds (mostly outside the 
barrack) this passive shield, acting as a further neutron moderator.
In particular, the neutron shield reduces by a factor 
larger than one order of magnitude the environmental thermal neutrons flux \cite{modlibra}.
The copper box is continuously maintained in HP Nitrogen atmosphere in slight overpressure with respect to the 
external environment; it is part of the 3-levels sealing system which prevents environmental air
reaching the detectors.
The DAMA/LIBRA-phase1 exposure (1.04 ton$\times$yr) has been collected during seven annual cycles \cite{modlibra,modlibra2,modlibra3}.
Considering also the former DAMA/NaI \cite{RNC,ijmd}, the total exposure collected
over 14 annual cycles is 1.33 ton$\times$yr. 
A clear modulation is present in the (2--6) keV {\it single-hit} events and fulfills all the requirements of the 
DM annual modulation signature; in particular, no modulation is observed either above 6 keV or in the
(2--6) keV {\it multiple-hits} events \cite{modlibra,modlibra2,modlibra3}.

The data provide a model
independent evidence of the presence of DM particles in the galactic halo at 9.3 $\sigma$ C.L.
on the basis of the investigated DM signature.
In particular, with the cumulative exposure the modulation
amplitude of the {\it single-hit} events in the (2--6) keV energy
interval, measured in NaI(Tl) target, is $(0.0112 \pm 0.0012)$
cpd/kg/keV; the measured phase is $(144 \pm 7)$ days (corresponding to May 24 $\pm$ 7 days) and the
measured period is $(0.998 \pm 0.002)$ yr, values well in agreement
with those expected for the DM particles.

Careful investigations
on absence of any significant systematics or side reaction able to
account for the measured modulation amplitude and to simultaneously satisfy
all the requirements of the signature
have been quantitatively carried out (see e.g. ref.
\cite{perflibra,modlibra,modlibra2,modlibra3,RNC,ijmd,mu,review}, and references therein);
none has been found or suggested by anyone over more than a decade. In particular, the cases
of the neutrons of whatever origin, and of the muons have been deeply investigated.

This paper summarizes in a simple and intuitive way why the neutrons, the muons and the
solar neutrinos cannot significantly contribute to 
the DAMA observed annual modulation signal; some of the 
already-published arguments \cite{perflibra,modlibra,modlibra2,modlibra3,RNC,ijmd,mu,review} 
are also recalled here. Afterwards, 
we demonstrate through few simple considerations the incorrectness of the claim reported in ref. \cite{davis}.

\section{Neutrons, muons and solar neutrinos at LNGS} 

In the following we discuss quantitatively the background induced by neutrons, muons and solar neutrinos in DAMA/LIBRA.

\subsection{The neutron flux at LNGS} 

The total flux of neutrons is given by the sum of
the thermal, the epithermal, and the fast components;
the latter can be written as:
\begin{equation}
\Phi^{(n)}_{fast} =  \Phi^{(n)}_{fiss,\alpha \rightarrow n} + \Phi^{(n)}_{\mu \rightarrow n} + \Phi^{(n)}_{\nu \rightarrow n};
\end{equation}
where:
i)   $\Phi^{(n)}_{fiss,\alpha \rightarrow n}$, the dominant component, is due to neutrons from fissioning elements and from $(\alpha,n)$ reactions 
     (neutron energy roughly from 1 to 10 MeV);
ii)  $\Phi^{(n)}_{\mu \rightarrow n}$ is due to neutrons generated by $\mu$ interactions (neutron energy distribution 
     with a long tail up to GeV's \cite{lvd_neut1,lvd_neut2,hime});
iii) $\Phi^{(n)}_{\nu \rightarrow n}$ is due to neutrons generated by solar neutrinos interactions (neutron energy roughly few MeV).
The possible yearly variation, if any, of each component $k$ can be pointed out by simply considering a cosine-like first term approximation:
\begin{equation}
\Phi^{(n)}_k =  \Phi^{(n)}_{0,k} \left( 1 + \eta_k cos \omega \left(t-t_k\right)\right),
\label{eq1}
\end{equation}
neglecting higher order harmonics; here $\omega=2\pi/(1$ year), $\eta_k$ is the relative modulation amplitude and 
$t_k$ the phase. 
The neutrons in the Gran Sasso caverns were measured many times by several authors \cite{bel85,rin88,neut1,ale89,neut2,arn99}.
The measurements are largely compatible among them; in particular,  
the flux of thermal neutron is $\simeq 1.08 \times 10^{-6}$ n cm$^{-2}$ s$^{-1}$ \cite{neut1},
the flux of epithermal neutron is $\simeq 2 \times 10^{-6}$ n cm$^{-2}$ s$^{-1}$ \cite{neut1}, and 
the flux of fast neutron is $\simeq 0.9  \times 10^{-7}$ n cm$^{-2}$ s$^{-1}$ \cite{neut2}.
In addition, the measured flux of neutrons with energy above 10 MeV, where the muon-induced neutrons mostly contribute,
is $\simeq 0.5 \times 10^{-9}$ n cm$^{-2}$ s$^{-1}$ \cite{neut1,arn99}.
It is worth noting that, at present, no compelling evidence of any time variation of the thermal, epithermal and fast neutrons is available (for details see
ref. \cite{modlibra,mu,review}). In the following calculations we cautiously adopt $\eta_k=0.1$ as done in ref. \cite{modlibra,mu,review};
thus, since $\eta_k \ll 0.1$ the derived values are cautelative upper limits.

We stress that these measurements regard all neutrons of whatever origin and, therefore,
even those induced by muons and by solar neutrinos. 
In particular, $\Phi^{(n)}_{0,\mu \rightarrow n}$ and $\Phi^{(n)}_{0,\nu \rightarrow n}$ are
less (or more correctly, much less) than the total flux of neutrons;
to estimate them, we adopt here the values reported in ref. \cite{davis} for 
the induced neutron rate production at LNGS\footnote{From ref. \cite{davis}:
i) the muon-induced neutron rate is $R_\mu \simeq 10^{-34} n V$ neutrons/s;
ii) the neutrino-induced neutron rate is $R_\nu \simeq  10^{-35} n V$ neutrons/s;
where:
$n$ is the number density of the target $\simeq 10^{29}$ m$^{-3}$ and
$V$ is the target volume.
Therefore, the induced neutron rate production per target volume unit is:
$r_\mu=R_\mu /V \approx 0.86$ n/day/m$^3$ for muon-induced interactions, and
$r_\nu=R_\nu /V \approx 0.09$ n/day/m$^3$ for solar neutrino-induced interactions.}:
$r_\mu \approx 0.86$ n/day/m$^3$ from muons\footnote{This value is well compatible with
the range reported in ref. \cite{mu}, considering the muon flux, the neutron yield and
the density of the rock at LNGS. For the case of lead target see in the following.} and
$r_\nu \approx 0.09$ n/day/m$^3$ from solar neutrinos.

A right calculation of the effective volume, $V_{eff}$, over which these neutrons are produced and reach the detectors
needs a simulation; for this purpose we can consider the DAMA/LIBRA setup
(about $51 \times 51 \times 25.4$ cm$^3$, that is its volume is $V_{LIBRA}=0.066$ m$^3$) at the 
center of a sphere of material where the neutrons are evenly and isotropically generated.

Taking into account the mean free path of the neutrons (cautiously assumed about 2.6 m as in ref. \cite{davis}), 
and considering a sphere with 10 m radius (about 4 times the neutrons mean free path), that is a volume
$V = 4200$ m$^3$, the mean probability that the induced neutrons reach the DAMA/LIBRA
setup can be calculated by the simulation: $P=0.016\%$; that is an effective volume $V_{eff} = V \times P = 0.7$ m$^3$.

The simulation\footnote{\label{ft3}It is worth noting that for the case of lead target
the neutron yield is \cite{lind09}:
$Y\simeq 1.3 \times 10^{-3}$ neutrons per muons per g/cm$^2$, and $r_\mu$ becomes $\simeq 29$ n/day/m$^3$;
considering that the total volume of the lead shield of DAMA/LIBRA is $V_{Pb} \simeq 1$ m$^3$,
and $P=4.5\%$ by the simulation, one can obtain an effective volume for the lead shield equal to
$V_{Pb} \times P = 0.045$ m$^3$. Thus, the corresponding neutron flux is $\simeq 6 \times 10^{-9}$ n cm$^{-2}$ s$^{-1}$;
however, this possible contribution to the DAMA annual modulation amplitude
is still negligible ($\ll 0.2\%$), even considering the cautious approches used in the following.} 
also provides $\Phi^{(n)}_{0,\mu \rightarrow n} \simeq 3 \times 10^{-9}$ n cm$^{-2}$ s$^{-1}$,
and $\Phi^{(n)}_{0,\nu \rightarrow n} \simeq 3 \times 10^{-10}$ n cm$^{-2}$ s$^{-1}$,
two orders of magnitude lower than the measured total one (given above).
This result is also compatible with the measurements at LNGS given above at neutron energy above 10 MeV \cite{neut1,arn99}, 
and with the expectations of ref. \cite{hime}, where a total muon-induced neutron flux 
$2.7 \times 10^{-9}$ n cm$^{-2}$ s$^{-1}$ has been calculated for LNGS.

A simple analytical calculation leads to the same conclusions; taking into account that every point around the DAMA/LIBRA setup
can be a source of neutrons induced by muons (neutrinos) with a rate $r_{\mu(\nu)}$ and cautiously assuming a mean free path $\lambda_0 = 2.6$ m,
one can write:
$$
\Phi^{(n)}_{0,\mu(\nu) \rightarrow n} = \int_V \frac{d\Phi(\vec{r})}{dV} d^3r = \int_V \frac{r_{\mu(\nu)}}{4\pi r^2} e^{-r/\lambda_0} d^3r = r_{\mu(\nu)} \int_0^\infty e^{-r/\lambda_0} dr = r_{\mu(\nu)}\lambda_0. 
$$
So that, $\Phi^{(n)}_{0,\mu \rightarrow n} = r_{\mu}\lambda_0 \simeq 2.6 \times 10^{-9}$ n cm$^{-2}$ s$^{-1}$ and 
and      $\Phi^{(n)}_{0,\nu \rightarrow n} = r_{\nu}\lambda_0 \simeq 2.6 \times 10^{-10}$ n cm$^{-2}$ s$^{-1}$, respectively,
confirming the results of the simulation.
The neutron fluxes are reported in Table \ref{table:tab12}.

\subsection{The muons at LNGS} 

The surviving muon flux, $\Phi^{(\mu)}$, has been measured in the deep
underground Gran Sasso Laboratory  at 3600 m w.e. depth 
by various experiments with very large exposures \cite{Mac97,LVD,borexino,borexino2,borexino3};
its mean value is $\Phi^{(\mu)}_{0} \simeq 20 $ muons m$^{-2}$d$^{-1}$ \cite{Mac97},
that is about a factor $10^6$ lower than that measured at sea level.
The measured average single muon energy at LNGS is
$[270 \pm 3 (stat) \pm 18(syst)]$ GeV \cite{Mac03,hime}.
A $\simeq$ 2\% yearly variation of the muon flux was firstly measured years ago by
MACRO; an extensive discussion about the muon flux variation along the year measured by the experiments at LNGS
has been reported in ref. \cite{mu}.
For the purpose of this paper we assume that:
\begin{equation}
\Phi^{(\mu)} =  \Phi^{(\mu)}_{0} \left( 1 + \eta^{(\mu)} cos \omega \left(t-t^{(\mu)}\right)\right);
\label{eq2}
\end{equation}
$t^{(\mu)}$ at LNGS location is at end of June (or later depending on each year; see e.g. ref. \cite{mu}
and references therein), and the relative modulation amplitude $\eta^{(\mu)}$ is 
$\simeq 0.0129\pm0.0007$ \cite{borexino2}.

\subsection{Solar neutrinos at LNGS} 

The total neutrino flux at LNGS is well established \cite{pdg}. Its time variability, experimentally pointed out 
-- still with modest C.L. -- by 
Super-Kamiokande (fiducial volume of 22.5 kton) \cite{kam1,kam2}, SNO (1000 tons) \cite{sno}, and Borexino (fiducial volume 100-150 tons) \cite{bor_mod}, is due to the different Sun-Earth distance along the year; so the relative modulation amplitude is twice the 
eccentricity of the Earth orbit and the phase is given by the perihelion. Thus, the total neutrino flux (from every source: pp, $^7Be$, $^8B$, pep, ...)
can be written as:
\begin{equation}
\Phi^{(\nu)} =  \Phi^{(\nu)}_{0} \left( 1 + \eta^{(\nu)} cos \omega \left(t-t^{(\nu)}\right)\right),
\label{eq3}
\end{equation}
where the phase, $t^{(\nu)}$, corresponds to Jan. 4th and the relative modulation amplitude $\eta^{(\nu)}$ is 0.03342.

\vspace{0.5cm}
The values of the parameters in eqs. \ref{eq1}, \ref{eq2} and \ref{eq3} are reported in Table \ref{table:tab12}.

\begin{sidewaystable}[hp]
\caption{Summary of the contributions to the total neutron flux at LNGS; the value, the relative modulation amplitude,
and the phase of each component is reported.  
It is also reported the counting rate in DAMA/LIBRA for {\it single-hit}
events, in the ($2-6$) keV energy region induced by neutrons, muons and solar neutrinos, detailed for each component.
The modulation amplitudes, $A_k$, are reported as well, while the last column shows the relative contribution 
to the annual modulation amplitude observed by DAMA/LIBRA, $S_m^{exp} \simeq 0.0112$ cpd/kg/keV \cite{modlibra3}.
As can be seen, they are all negligible and they cannot give any significant contribution to
the observed modulation amplitude. In addition, neutrons, muons and solar neutrinos are not a competing background 
when the DM annual modulation signature is investigated since in no case they can mimic this signature. See text.}
\begin{center}
\resizebox{0.98\textwidth}{!}{
\begin{tabular}{|ll|ccc|ccc|c|}
\hline 
\multicolumn{2}{|c|} {Source} & $\Phi^{(n)}_{0,k}$                & $\eta_k$                                    & $t_k$ & $ R_{0,k}$   & & $A_k = R_{0,k} \eta_k $        & $A_k/S_m^{exp}$ \\
 &                     & (neutrons cm$^{-2}$ s$^{-1}$)            &                                             &       & (cpd/kg/keV) & & (cpd/kg/keV) & \\
\hline 
 & thermal n & $1.08 \times 10^{-6}$ \cite{neut1}    & $\simeq 0$                                  & --    & $<8 \times 10^{-6}$ & \cite{modlibra,mu,review} & $\ll 8 \times 10^{-7}$ & $\ll 7 \times 10^{-5}$ \\
 & ($10^{-2}-10^{-1}$ eV) &                          & however $\ll 0.1$ \cite{modlibra,mu,review} &       &   &   & & \\
 SLOW & & & & & & & & \\
 neutrons & epithermal n & $2 \times 10^{-6}$ \cite{neut1}          & $\simeq 0$                                  & --    &   $<3 \times 10^{-3}$ & \cite{modlibra,mu,review} & $\ll 3 \times 10^{-4}$ & $\ll 0.03$ \\
 & (eV-keV)     &                                          & however $\ll 0.1$ \cite{modlibra,mu,review} &       &   &   & & \\
\hline 
 & fission, $(\alpha,n) \rightarrow$ n  & $\simeq 0.9 \times 10^{-7}$ \cite{neut2} & $\simeq 0$                                  & --  & $< 6 \times 10^{-4}$ & \cite{modlibra,mu,review} & $\ll 6 \times 10^{-5}$ & $\ll 5 \times 10^{-3}$ \\
 & (1-10 MeV)               &                                          & however $\ll 0.1$ \cite{modlibra,mu,review} &     &   &   & & \\
 & & & & & & & & \\
     & $\mu \rightarrow $ n from rock    & $\simeq 3 \times 10^{-9}$                & 0.0129 \cite{borexino2}                     & end of June \cite{borexino2,mu,review} & $\ll 7 \times 10^{-4}$ & (see text and               & $\ll 9 \times 10^{-6}$ & $\ll 8 \times 10^{-4}$ \\
 FAST  & ($> 10$ MeV)             & (see text and ref. \cite{hime})         &                                             &                                       &                        & \cite{modlibra,mu,review})  &   & \\
 neutrons & & & & & & & & \\
     & $\mu \rightarrow $ n from Pb shield & $\simeq 6 \times 10^{-9}$                & 0.0129 \cite{borexino2}                     & end of June \cite{borexino2,mu,review} & $\ll 1.4 \times 10^{-3}$ & (see text and               & $\ll 2 \times 10^{-5}$ & $\ll 1.6 \times 10^{-3}$ \\
     & ($> 10$ MeV)             & (see footnote \ref{ft3})         &                                             &                                       &                        & footnote \ref{ft3})  &   & \\
 & & & & & & & & \\
 & $\nu \rightarrow $ n     & $\simeq 3 \times 10^{-10}$ (see text)    & 0.03342  $^*$                               & Jan. 4th $^*$ & $\ll 7 \times 10^{-5}$ & (see text) & $\ll 2 \times 10^{-6}$ & $\ll 2 \times 10^{-4}$ \\
 & (few MeV)                & & &   &   &   &  & \\
\hline 
\multicolumn{2}{|c|} {direct $\mu$}       & $\Phi^{(\mu)}_{0} \simeq 20$ $\mu$ m$^{-2}$d$^{-1}$ \cite{Mac97}   & 0.0129 \cite{borexino2} & end of June \cite{borexino2,mu,review} & $\simeq 10^{-7}$ & \cite{modlibra,mu,review} & $\simeq 10^{-9}$ & $\simeq 10^{-7}$ \\
 & & & & & & & & \\
\multicolumn{2}{|c|} {direct $\nu$}       & $\Phi^{(\nu)}_{0} \simeq 6 \times 10^{10}$ $\nu$ cm$^{-2}$s$^{-1}$ \cite{pdg} & 0.03342  $^*$          & Jan. 4th $^*$                         & $\simeq 10^{-5}$ & \cite{multiton} & $3 \times 10^{-7}$ & $3 \times 10^{-5}$ \\
\hline 
\end{tabular}}
\end{center}
\label{table:tab12}
{\footnotesize $^*$ The annual modulation of solar neutrino is due to the different Sun-Earth distance along the year; so the relative modulation amplitude is twice the 
eccentricity of the Earth orbit and the phase is given by the perihelion.}
\end{sidewaystable}

\section{Counting rate in DAMA/LIBRA induced by neutrons, muons and solar neutrinos} 

The counting rate of the DAMA/LIBRA detectors for {\it single-hit}
events, in the ($2-6$) keV energy region due to neutrons, muons and solar neutrinos
can be written as the sum of seven contributions due to:
\begin{enumerate}
\item thermal neutrons ($R_{thermal \, n}$);
\item epithermal neutrons ($R_{epith. \, n}$);
\item fast neutrons due to fissioning elements and to $(\alpha,n)$ reactions ($R_{fiss,\alpha \rightarrow n}$);
\item fast neutrons generated by $\mu$ interactions ($R_{\mu \rightarrow n}$);
\item fast neutrons generated by solar neutrinos interactions ($R_{\nu \rightarrow n}$);
\item direct interactions of muons ($R_{direct \; \mu}$);
\item direct interactions of solar neutrinos ($R_{direct \; \nu}$).
\end{enumerate}
The possible yearly modulation, if any, of each component $k$ can be described as:
\begin{equation}
R_k =  R_{0,k} \left( 1 + \eta_k cos \omega \left(t-t_k\right)\right).
\end{equation}
The modulation amplitude of the component $k$ is $A_k = R_{0,k} \eta_k $.
The known parameters are reported in Table \ref{table:tab12}.

In particular, the $R_{0,k}$ from neutrons and from muons were calculated in ref. \cite{modlibra,mu,review},
while the one from ``direct'' neutrino interactions was calculated in ref. \cite{multiton}.
As concern the $R_{0,k}$ from muon-induced and neutrino-induced neutrons, they can be calculated by the simplified model used above.
For simplicity, we do not take into account e.g. the further reducing effect of the neutron shield of DAMA/LIBRA.
One has:
1) neutrons directly induced in the DAMA/LIBRA setup = $r_{\mu(\nu)} V_{LIBRA} = 0.06$ (0.006) neutrons/day for muon- (neutrino-) induced events;
2) neutrons induced in the surrounding sphere $= r_{\mu(\nu)} V_{eff} = 0.58$ (0.06) neutrons/day for muon- (neutrino-) induced events.

Therefore, even for detection efficiency equal to 1, the rate from muon-induced neutrons in DAMA/LIBRA cannot exceed 0.64 cpd;
more likely the rate is $\ll 0.64$ cpd. 
Maximizing the effect, and even assuming that such counts
are only {\it single-hit} and in the 2-6 keV energy region, the counting rate $\ll \frac{0.64 cpd}{4 keV 242.5 kg} \simeq 7 \times 10^{-4}$ cpd/kg/keV
can be obtained (to be compared with the {\it single-hit} total rate of the DAMA detectors that is around 1 cpd/kg/keV \cite{perflibra}).
One order of magnitude less is from neutrino-induced neutrons. 

We note that the modulation amplitude for the case of neutrons induced by muons (see Table \ref{table:tab12})
is well compatible with the value estimated using different approach \cite{modlibra,mu,review}: $\ll (3-24) \times 10^{-6}$ cpd/kg/keV.

All these values are reported in Table \ref{table:tab12}.
The last column shows for each component the relative contribution 
to the annual modulation amplitude observed by DAMA/LIBRA.
As can be seen, they are all negligible and they cannot give any significant contribution to
the observed modulation amplitude.

In addition, it is worth noting that neutrons, muons and solar neutrinos are not a competing background when the DM annual modulation signature is investigated since 
in no case they can mimic this signature; in fact,
some of its peculiar requirements fail, such as
they would induce e.g. variations in all the energy spectrum,
variation in the {\it multiple-hit} events,... These latter ones were consistently not observed in DAMA data \cite{modlibra,modlibra2,modlibra3,mu,review}.

\section{Four simple arguments against the claim of ref. \cite{davis}}

The arguments summarized above clearly demonstrate that the claim in ref. \cite{davis} is incorrect.
In this Section we report in addition just few simple underlying considerations
that would already have been enough to clearly demonstrate the incorrectness of the claim in ref. \cite{davis}.

The paper of ref. \cite{davis} reports about a fit on the residuals of DAMA/LIBRA annual
modulation result \cite{perflibra,modlibra,modlibra2,modlibra3,mu,review}, using the following function:
\begin{equation}
A_{\mu+\nu} = A_\nu cos(\omega(t-\phi_\nu)) + A_\mu cos(\omega(t-\phi_\mu));
\end{equation}
where the two contributions are claimed due to neutrons produced by solar neutrino and muon interactions, respectively.
The frequency and phases are constrained to: $\omega=2\pi/(1$ year), $\phi_\nu \simeq$ January 4th, and $\phi_\mu \simeq$ end of June (see also ref. \cite{mu,review}),
while the two modulation amplitudes, $A_\nu$ and $A_\mu$, are considered as free parameters of the fit.
This ``mathematical'' exercise produces two ``big'' modulation amplitudes since a sort of cancellation occurs
between the two effects, having quasi-opposite phases. 

But this ``mathematical'' exercise does not represent a physical possibility for many reasons, as shown in the following.

\vspace{0.5cm}
\noindent {\bf 1. Neutrons from whatever origin (even from muons and from solar neutrinos) cannot give any relevant contribution to
the DAMA modulation effect.}

It has been demonstrated (see e.g. \cite{modlibra,mu,review})
that the neutrons -- independently of their origin -- cannot give any relevant contribution to
the DAMA modulation effect.
In fact the neutrons, surviving the heavy shield against them, can be quantitatively studied in
various ways in the DAMA experiment \cite{modlibra,mu,review}; even when hypothetically and cautiously assuming 
a 10\% modulation (of whatever origin) of neutrons flux with the same phase and
period as for the DM case, the corresponding modulation amplitude is 
two/three orders of magnitude lower than the DAMA observed modulation amplitude \cite{modlibra,modlibra2,modlibra3,mu,review} (see 
Table \ref{table:tab12}).

In conclusion, considering that:
i) the flux of neutrons induced by muons and by solar neutrinos is less (or more correctly, much less) than the total flux of neutrons (see Table \ref{table:tab12}),
ii) the relative modulation amplitude of neutrons induced by muons is about $1.3\%$, as shown in Table \ref{table:tab12} (and thus $<10\%$ assumed above),
iii) the relative modulation amplitude of neutrons induced by solar neutrinos is 0.03342 from the eccentricity of the Earth orbit (and thus $<10\%$ assumed above),
any hypothetical contribution of neutrons -- and in particular those from muons and solar neutrinos --
to the DAMA annual modulation effect is absolutely negligible.

In addition, as mentioned above, it is worth noting that neutrons are not a competing background when the DM annual modulation signature is investigated since 
in no case they can mimic this signature; in fact,
some of its peculiar requirements fail, such as
the neutrons would induce e.g. variations in all the energy spectrum,
variation in the {\it multiple-hit} events,... These latter ones were consistently not observed in DAMA data \cite{modlibra,modlibra2,modlibra3,mu,review}.

\vspace{0.5cm}
\noindent {\bf 2. Results of the fit reported in ref. \cite{davis} lead to erroneous claims.}

The fitting procedure, reported in ref. \cite{davis}, on the {\it single-hit} experimental residuals measured
by DAMA leads to the absurd fact that the fitted modulation
amplitude $A_\mu$ is of the same amount as $A_\nu$.

In particular, the fitting procedure in ref. \cite{davis} yields to $A_\nu \simeq 0.039$ cpd/kg/keV and
$A_\mu \simeq 0.047$ cpd/kg/keV; using the relative modulation amplitudes of solar neutrinos
and muons given above, 
one can easily determine in such hypothesis the respective contributions to the {\it single-hit} total rate of the DAMA/LIBRA detectors.
They would be 0.039/0.03342 = 1.17 cpd/kg/keV and 0.047/0.0129 = 3.6 cpd/kg/keV, respectively.
Thus, that fit produces two ``big'' contributions, much larger (even orders of magnitude) than
their correct estimates (see above and Table \ref{table:tab12}) and much larger than the measurements as well. We remind that the ${\it single-hit}$ 
total rate of the DAMA detectors is around 1 cpd/kg/keV  (see e.g. \cite{perflibra}).

\vspace{0.5cm}
\noindent {\bf 3. DAMA vs solar neutrinos: may DAMA/LIBRA compete with the multi-hundred-ton experiments 
for the detection of $^8$B solar neutrinos?}

As shown in Table \ref{table:tab12}, 
the rate for ``direct'' interactions of solar neutrinos 
on NaI(Tl) is around $10^{-5}$ cpd/kg/keV \cite{multiton} at low energy,
many orders of magnitude lower than the measured total {\it single-hit} rate in DAMA.

The author of ref. \cite{davis} claims that the ``indirect'' (neutrons induced just by $^8$B solar neutrinos) 
solar neutrino event rate would be 1.17 cpd/kg/keV (see above).
That is the ``indirect'' contribution is claimed to be many orders of magnitude larger that the ``direct'' one; may DAMA/LIBRA compete with multi-hundred-ton experiments 
for the detection of $^8$B neutrinos and even be able to see its annual variation!?

\vspace{0.5cm}
\noindent {\bf 4. Ref. \cite{davis} reports three (at least) orders of magnitude wrong estimates of the counting rate in DAMA/LIBRA 
due to neutrons from neutrinos and muons.}

Taking into account the mean free path (cautiously about 2.6 m) of the muon-induced neutrons, Davis estimates:
``the effective volume over which these neutrons are produced and still reach the detector to be
... 450 m$^3$''.

This effective volume is wrong by several orders of magnitude, because
the author of ref. \cite{davis} does not take into account e.g. the geometrical efficiency in DAMA/LIBRA
for the detection of those produced neutrons.
This effective volume has been properly calculated above: $V_{eff} \simeq 0.7$ m$^3$.

In addition, as reported in Table \ref{table:tab12},
the induced modulation amplitudes from neutrons induced by muons and by neutrinos are
$\ll 9 \times 10^{-6}$ cpd/kg/keV ($\ll 2 \times 10^{-5}$ cpd/kg/keV for neutrons produced in the lead shield)
and $\ll 2 \times 10^{-6}$ cpd/kg/keV,
respectively. They are $\ll 0.1\%$ of the measured modulation amplitude by DAMA/LIBRA.
These upper limits are orders of magnitude lower than the $A_\mu$ and $A_\nu$ values claimed in ref. \cite{davis}. 

\vspace{0.7cm}
In conclusion, already just considering the arguments given above the claim of ref. \cite{davis} is unfounded.

\section{Conclusions} 

This paper further summarizes in a simple and intuitive way why the neutrons,
the muons and the solar neutrinos cannot significantly contribute to the DAMA annual
modulation results. 
In addition, neutrons, muons and solar neutrinos are not a competing background 
when the DM annual modulation signature is investigated since in no case they can mimic this signature.
A number of arguments have already been presented in 
individual papers by DAMA collaboration. Few simple considerations are summarized 
which demonstrate the incorrectness of the claim reported in ref. \cite{davis}.

\end{document}